\documentstyle[12pt]{article}
\newcommand{\rf}[1]{(\ref{#1})}
\newcommand{\bea}{\begin{eqnarray}}
\newcommand{\eea}{\end{eqnarray}}
\newcommand{\g}{\gamma}
\renewcommand{\l}{\lambda}
\renewcommand{\b}{\beta}

\newcommand{\n}{\nu}
\newcommand{\m}{\mu}

\newcommand{\sg}{\sigma}

\newcommand{\ra}{\rangle}
\newcommand{\la}{\langle}

\def\void{}
\def\labelmark{}

\newenvironment{formula}[1]{\def\labelname{#1}
\ifx\void\labelname\def\junk{\begin{displaymath}}
\else\def\junk{\begin{equation}\label{\labelname}}\fi\junk}%
{\ifx\void\labelname\def\junk{\end{displaymath}}
\else\def\junk{\end{equation}}\fi\junk\labelmark\def\labelname{}}

{\ifx\void\labelname\def\junk{\end{array}\end{displaymath}}
\else\def\junk{\end{array}\right.\end{equation}}
\fi\junk\labelmark\def\labelname{}\def\junk{}
}

\newcommand{\beq}{\begin{formula}}
\newcommand{\eeq}{\end{formula}}
\newcommand{\beqv}{\begin{formula}{}}

\begin{document}
\topmargin 0pt
\oddsidemargin 5mm
\headheight 0pt
\headsep 0pt
\topskip 9mm

\addtolength{\baselineskip}{0.20\baselineskip}
\hfill    NBI-HE-92-31

\hfill May 1992
\begin{center}

\vspace{36pt}
{\large \bf

3d quantum gravity coupled to matter}

\vspace{36pt}

{\sl J. Ambj\o rn }

\vspace{12pt}

 The Niels Bohr Institute\\
Blegdamsvej 17, DK-2100 Copenhagen \O , Denmark\\

\vspace{12pt}

{\sl Z. Burda and J. Jurkiewicz}{\footnote {Supported by the
KBN grant no. 2 0053 91 01}}

\vspace{12pt}

Inst. of Phys., Jagellonian University., \\
ul. Reymonta 4, PL-30 059, Krak\'{o}w~16, Poland

\vspace{12pt}

{\sl C.F. Kristjansen}

\vspace{12pt}

 The Niels Bohr Institute\\
Blegdamsvej 17, DK-2100 Copenhagen \O , Denmark\\

\vfill
\end{center}

\vspace{24pt}

\begin{center}
{\bf Abstract}
\end{center}

\vspace{12pt}

\noindent
We investigate the phase structure of three-dimensional quantum gravity
coupled to an Ising spin system by means of numerical simulations.
The quantum gravity part is modelled by the summation over random
simplicial manifolds, and the Ising spins are located in the center
of the tetrahedra, which constitute the building blocks of the piecewise
linear manifold. We find that the coupling between spin and geometry
is weak away from the critical point of the Ising model. At the
critical point there is clear coupling, which however does not seem
to change the first order transition between the ``hot'' and ``cold''
phase of three dimensional simplicial quantum gravity observed earlier.

\vspace{24pt}

\vfill

\newpage

\section{Introduction}

In the recent years there has been a remarkable progress in our understanding
of two-dimensional quantum gravity coupled to matter.  Although it is
still somewhat unclear what kind of theory we deal with when we
consider pure 2d quantum gravity (how is the ``average'' intrinsic geometry
to be characterized etc.), we get definite answers, if we
ask for the modification of critical exponents of
conformal field theories when they are  coupled covariantly
to 2d quantum gravity. From the point of view of string theory it is
important to study the two-dimensional models, but if we want to
consider possible theories of quantum gravity outside the context of
string theories (and there are good reasons for taking such a point of view)
they are to be considered only as toy models for higher dimensional
gravity. In this letter we attempt to take a first step
in the direction of the study of higher dimensional gravity coupled
to matter. We will confine ourselves to three dimensions.
There are several reasons for this. Firstly we do not have available the
same powerful analytical methods as in two dimensions
and we have to rely on
numerical simulations. These are quite a lot easier in three dimensions
than in four dimensions.
In addition we expect three dimensional gravity to be placed in between
the solvable two-dimensional case and four-dimensional gravity in the
following sense: Three-dimensional (classical) gravity has
no dynamical degrees of freedom by itself, as is also the case for
two-dimensional gravity. But as for two-dimensional gravity this does
not imply that it cannot couple in a non-trivial way to the matter fields.
On the other hand the theory shares with the four-dimensional theory
the feature that the Einstein-Hilbert term in the action is non-trivial,
not renormalizable and (if not regularized) unbounded from below.
{}From this point of view we might get important hints
which can be used in four dimensions from a theory, where we can hopefully
use some of the intuition we have now gathered in the two-dimensional
studies. Furthermore it is much easier to find non-trivial statistical
models in three- than in four dimensions. In fact the simplest example is
the Ising model which we are going to use.

\section{The model}

Three-dimensional quantum gravity modelled on random triangulations
was introduced in refs. \cite{adj,sakuri,gross} and numerical simulations
first performed in refs. \cite{av,bk,am,av1,abkv}. Let us briefly
describe the model and summarize the results.

The continuum action of Einstein-Hilbert gravity in three dimensions
can be written as
\beq{*1}
S[g] =\l \int d^3 \xi \sqrt{g} - \frac{1}{16\pi G } \int d^3\xi \sqrt{g}\; R.
\eeq
In the dynamical triangulated approach the functional integral over metrics
is replaced by a summation over all possible triangulations. Here, as
in two dimensions, it is important that we restrict ourselves to a
fixed topology (which we take to be that of $S^3$). The building blocks
which we glue together in order to form the piecewise linear manifolds
are regular tetrahedra. Consequently we will get (see for instance \cite{av1}
for a careful discussion)
\bea
\int d^3 \xi \sqrt{g} &\sim & N_3    \label{*2} \\
\int d^3 \xi \sqrt{g} \; R & \sim & cN_1 - 6 N_3  \label{*3}
\eea
where $N_0,...,N_3$ denote the number of vertices, links, triangles and
tetrahedra which constitute the simplicial manifold. The number
$c = 2\pi/\arccos (1/3)$ is chosen such that $R =0$ corresponds to flat
three-dimensional space. This means that the discretized action can
be written as
\beq{*4}
S[T] = k_3 N_3 -k_1 N_1
\eeq
and the recipe for going from the continuum functional to the discretized
one will be:
\beq{*5}
\int {\cal D} [g_{\m\n}]e^{-S[g]} \to \sum_T e^{-S[T]}.
\eeq
The result of the numerical simulations is as follows:
For small $k_1$ (i.e. for large bare gravitational coupling constant)
 the system seems to be in a phase of large Hausdorff dimension
(denoted the ``hot'' phase). Even for large systems consisting of
28.000 tetrahedra the linear extension (the average geodesic distance)
of the system is small and increases only slowly with volume. For
$k_1 \approx 4.0$ there is a phase transition to a state where the
system is quite extended. The extension grows linear with the volume,
showing that the Hausdorff dimension is one. This phase is probably a
lattice artifact which signals the dominance of the conformal
mode of gravity for a small gravitational constant.
The transition between the two phases is  a first
order transition since  pronounced hysteresis is observed. We conclude that
if we restrict ourselves to actions of the form \rf{*4} it seems not possible
to define a continuum limit of the lattice model in the usual sense, with
a divergent correlation length. This is not an undesirable
situation since the presence of a divergent correlation length would
force us to identify at least one massless field in three-dimensional
quantum gravity. But we know such fields are not present in classical
three-dimensional gravity.

\vspace{12pt}

Let us couple the above defined model to Ising spins. The coupling
is done in the same way as for discretized two-dimensional quantum
gravity, where we know the dynamical triangulated model coupled to
Ising spins leads to critical exponents which agree with the ones
calculated using continuum formalism. To each tetrahedron $i$ we associate
an Ising spin $\sg_i$ and the partition function for the combined system
is
\beq{*6}
Z(\b,k_1,k_3) = \sum_{N_3} e^{-k_3N_3} \sum_{T \sim N_3} \sum_{[\sg]}
e^{k_1 N_1} e^{\b \sum_{<i,j>}(\delta_{\sg_i \sg_j} - 1)}.
\eeq
In this formula $T \sim N_3$ signifies the summation over all piecewise
linear manifolds which can be formed by gluing $N_3$ (regular) tetrahedra
together such that the topology is that of $S^3$. $\sum_{[\sg]}$ means
the summation over all spin configurations while $\sum_{<i,j>}$ stands for
the summation over all neighbour pairs of tetrahedra.  One annoying aspect
of the above formalism is that we are forced to perform a grand canonical
simulation where $N_3$ is not fixed. The reason is that we (contrary to
two dimensions) have no ergodic updating algorithm which preserves
the volume $N_3$. In practise it is however possible to perform the
measurements at a fixed $N_3$ and the important coupling
constants will then be $\b$ and $k_1$. We refer to \cite{av1} for
a detailed discussion. The spin updating is performed by  the single
cluster variant of the Swendsen-Wang algorithm developed by Wolff \cite{wolf}.
The cluster updating algorithms have been successfully applied to
the Ising model coupled to 2d gravity \cite{jj,rk,bj,bj1}
and to the ordinary three dimensional Ising model \cite{bghp}.

\section{ Numerical results}

As mentioned above three-dimensional simplicial quantum gravity has
two phases depending on the value of $k_1$.
The first statement we can make is that this is unchanged by the
coupling to Ising spin.

In the ``hot'' phase
($k_1 \le 4.0$) where the Hausdorff dimension is large
the numerical value of the magnetization
\beq{*7}
 | \sg |\equiv \frac{1}{N_3} \left|\sum_{i=1}^{N_3} \sg_i \right|
\eeq
is shown in fig.1 as a function of $\b$.
We see a clear signal indicating a phase transition
from a disordered phase (small $\b$)
where $ |\sg | \approx 0$  to an ordered phase (large $\b$) where
$| \sg | \approx 1$. The transition becomes  sharper
with increased volume $N_3$ and seems to be a second order transition.
This situation is contrasted by the magnetization curve in the ``cold'' phase
shown in fig.2. Here is only a gradual cross over to $| \sg | \approx
1$ for large $\b$, and the cross over is weakened  for increased
volume $N_3$. The situation is precisely as one
would expect in the case of a one-dimensional system where there
is no spontaneous magnetization. We conclude that the
Hausdorff dimension $d_H \approx 1$ measured in the pure gravity case seems
to reflect correctly the dimension relevant for coupling to matter.

In the $k_1-\b$ plane we have the phase-diagram shown in fig.3.
If $\b$ is away from the critical value $\b_c(k_1)$ (which has only
a weak dependence on $k_1$) the coupling between the fluctuations
in geometry and spin seems weak and of course it vanishes  in the
limits $\b \to \infty$ and $\b \to 0$. In these limits we therefore have
a strong first order transition between the ``hot'' and the ``cold'' phase
of three-dimensional quantum gravity, precisely as is  the case in
the absence of spins \cite{abkv}.
In the ``hot'' phase, where the Ising system has a second order transition, we
have seen an increased coupling between geometry and spins when we
approach the critical $\b_c(k_1)$.  This is shown in fig.4 where we
plot the average curvature $\la R \ra$ as a function of $\b$. A clear
peak is seen at $\b_c$. This enhanced coupling between geometry and spins
at the critical point is qualitatively in agreement
with the 2d results, where we  have a change in the string susceptibility
exponent $\g_{string}$ (not to be confused with the magnetic susceptibility
exponent $\g_{mag}$) from the pure gravity value $-1/2$ to $-1/3$,
precisely when $\b=\b_c$.
Unfortunately it is not clear that the entropy exponent analogous to
$\g_{string}$ exists in the hot phase of three-dimensional quantum
gravity (\cite{av,bk,av1,abkv}) so we have no obvious exponent with
which we can compare the effect of the spin coupling, but the enhanced
coupling between spin and geometry leaves open the possibility that the
transition between the ``hot'' and ``cold'' phase changes from a
first order to a second order transition.
We have looked for hysteresis when changing $k_1$ and
adjusting $\b$ to the critical value $\b_c(k_1)$. While the hysteresis
is indeed weaker when measured this way, we still see a clear hysteresis
(fig.5) and we conclude that there is never a second order transition
in geometry.

Let us make the following remark concerning the determination of the
phase diagram shown in fig.3: Due  to the strong hysteresis it is
somewhat ambiguous. We have used the following procedure: Well inside
the ``hot'' phase the system follows a unique path when changing
 $k_1$ and keeping $\b$ fixed as illustrated in fig.5.
The precise location depends on the
value of $\b$. We have extrapolated these paths until they intersect
the  parts of the hysteresis curves which correspond to the ``cold''
phase.

In fig.6 we have shown the spin-spin correlation as a function of
geodesic distance. To be precise there are two obvious candidates
for geodesic distances (see \cite{aj} for a discussion in the context
of four-dimensional simplicial quantum gravity). We can define the
geodesic distance $d_1$ between two vertices as the length of the
shortest path along links
connecting the two vertices. Alternatively we could have defined the geodesic
distance $d_2$ between two tetrahedra as the length of the
shortest path connecting the two tetrahedra, moving from center
to center in neighbour tetrahedra which have a triangle in common.
The first definition is clearly much closer to the ``correct'' definition
obtained by considering the manifolds as piecewise linear, with the
curvature attached to the links. The other definition corresponds to
moving along links in the {\it dual} graph, which is a $\phi^4$ graph.
For a single manifold the two distances can differ a lot, but when
an ensemble average is taken it seems as if one can consider them as
proportional. A similar result is true in four-dimensional gravity
(\cite{aj}). Clearly $d_2$ is most convenient for our purpose and starting
from a given tetrahedron $i_0$ we define the volume $V_2(r)$
inside a ball of
$d_2$-geodesic  radius $r$ around $i_0$ as the number of
tetrahedra within this
distance. Further the differential volume is $dV_2(r)\equiv
V_2(r) - V_2(r-1)$. We can now define a spin-spin correlation function as
\beq{*8}
g(r) \equiv \left\la \frac{1}{dV_2(r)} \sum_{i \in dV_2(r)} \sg_i \sg_{i_0}
\right\ra .
\eeq
An alternative correlation function would be
\beq{*9}
G(r) \equiv \left\la  \sum_{i \in dV_2(r)} \sg_i \sg_{i_0} \right\ra.
\eeq
which is related to the magnetic susceptibility $\chi (\b)$ by
\beq{*9a}
\sum_{r} G(r) = \chi(\b) \sim |\b-\b_c(k_1)|^{-\g}~~~~{\rm for}
{}~~~\b \to \b_c(k_1)
\eeq
In fig.6 we have shown $g(r)$ for two
different values of $\b$. In principle one can extract the mass gap
$m(\b)$ from the exponential fall off of $g(r)$
and in this way determine the critical
exponent $\n$ defined by $m(\b)= |\b-\b_c|^\n$. This seems however difficult
to do in a reliable way, in accordance with the experience in two-dimensional
quantum gravity coupled to Ising spins, and it is maybe understandable if
one keeps in mind that not only is a precise determination of
$\b_c$ needed in order to extract $\n$. In addition
our data are folded into the distribution $dV_2(r)$ which determines the
Hausdorff dimension, a quantity which by itself
is very difficult to measure in a reliable way.

In the same way we can construct
$\chi(\b)$ from $G(r)$, but it does not lead to a precise determination
of the critical exponent $\g$ (again in agreement with the experience
from two dimensions).

\section{Discussion}

We have shown that the phase structure of three-dimensional
simplicial quantum gravity, as described in \cite{abkv}, is
not modified by the presence of matter, at least in the simplest
case of coupling to Ising spins. In the so-called ``hot'' phase
the Ising spin system has a second order transition, while it has
no transition in the ``cold'' phase in agreement with the effective
one-dimensional nature of this phase. The existence of two phases in
three-dimensional gravity is caused by the Einstein-Hilbert term
in the action. In two dimensions this term is absent (for a fixed topology)
and we have only one phase of pure gravity. This phase seems to have most
in common with the ``hot'' phase of three-dimensional gravity and
it is natural to expect that the critical properties of the matter
theories covariantly coupled to 3d gravity could be
changed in a non-trivial way when we are in this ``hot'' phase,
simply by analogy with the two-dimensional models.

One way to investigate the possible non-trivial scaling
of the Ising model in the ``hot'' phase  is by means of finite
size scaling. In two-dimensional gravity this approach seems to work
somewhat better than the direct attempts to measure $\n$ and $\g$ mentioned
above. The disadvantage of the method is that it only gives us certain
combinations of the exponents.

If a given thermodynamic function $F$ has a critical
behaviour
\beq{*10}
F(\b) \sim (\b-\b_c)^{-x}
\eeq
one expects in ordinary flat space a finite size dependence of the
form
\beq{*11}
F(\b,L) = L^{\frac{x}{\n}} f( |\b-\b_c| L^{1/\n})
\eeq
where $L$ denotes the linear size of the system and
the exponent $\n$ is determined by the divergence of the
correlation length $\xi(\b)$:
\beq{*12}
\xi(\b) \sim |\b-\b_c|^{-\n}.
\eeq
By a measurement of $F(\b_c,L) \sim f(0) L^{\frac{x}{\n }}$ as a function
of $L$ we can determine the combination $x/\n $, while  measurements
away from $\b_c$
would give us $x$ directly for sufficiently large $L$.

If we want to use these formulae for systems coupled to quantum gravity
we must identify the divergent correlation length $\xi$ in terms of
geodesic distances, as was already discussed in the last section. We
further have to identify the linear extension $L$.  If the system has
a finite Hausdorff dimension $d_H$ it is tempting to define
\beq{*13}
L \sim N_3^{1/d_H}.
\eeq
These ideas have been used with some success in 2d-quantum gravity
\cite{rk,bj,bj1}, but could be spoiled if the  Hausdorff dimension
is infinite. Since the $d_H$
from (very tentative) direct measurements of $V_2(r)$ seems
large in the ``hot''
phase {\it it is tempting to conjecture that the critical exponents
of the Ising model in the ``hot'' phase take their mean-field values}.
A reliable determination of the various exponents along the lines
discussed above requires a considerable amount of computer time,
since it is already quite demanding in two dimensions, but we hope
to be able to address the question in a future publication.

\vspace{12pt}
\noindent
Note added: While completing this article we received a paper by
Baillie \cite{baillie},
who investigates the same system. There seems to be little
overlap with our work since he did not explore the phase structure
in $k_1-\b$-plane. In fact this is not possible with the size of systems
he uses.

\newpage

\noindent {\large \bf Figure Captions}

\vspace{12pt}

\begin{itemize}

\item[Fig.1] The magnetization $|\sg|$ (defined by \rf{*7})
as a function of $\b$ in the ``hot'' phase for
$N_3=$ 4000 (triangles) and $N_3=$ 10000 (circles).

\item[Fig.2] The magnetization $|\sg|$ (defined by \rf{*7})
as a function of $\b$ in the ``cold'' phase for
$N_3=$ 4000 (triangles) and $N_3=$ 10000 (circles).

\item[Fig.3] The phase diagram for 3d quantum gravity coupled to matter.
Filled circles are results obtained for $N_3=$ 10000.

\item[Fig.4] The average curvature $\la R \ra$ as a function of $\b$
for $N_3=$ 4000 (circles) and $N_3=$ 10000 (squares).
The position of the peak coincides with the value of $\b_c$ determined
from the magnetization curve.

\item[Fig.5] The hysteresis curve for pure gravity (triangles) and
in the case where the Ising spin system is critical i.e. where it couples
in a maximal way to gravity (circles). $N_3=$ 10000.

\item[Fig.6] The spin-spin correlation function $g(r)$ (defined by \rf{*8})
 as  a function of the geodesic
distance $r$ for $\b=$ 0.5 (full drawn curve) and $\b=$ 0.8 (dotted
curve),
$k_1=$ 3.7 and $N_3=$ 4000. The best estimate of the critical value
of $\b$ is: $\b_c=$ 0.85

\end{itemize}

\end{document}